# AVANÇOS NA CARACTERIZAÇÃO DAS ESTRUTURAS GEOLÓGICAS EM SUBSUPERFÍCIE DA PROVINCIA URANÍFERA LAGOA REAL (BA) A PARTIR DE DADOS AEROGEOFÍSICOS

**Suze Nei Pereira GUIMARÃES & Valiya Mannathal HAMZA**

Observatório Nacional, Ministério da Ciência e Tecnologia, ON/MCT. Rua General José Cristino, 77 – Bairro Imperial de São Cristóvão. CEP 20921-400. Rio de Janeiro, RJ. Endereços eletrônicos: hamza@on.br; suze@on.br

**RESUMO** – O presente trabalho traz os resultados de uma reavaliação interpretativa dos dados aerogeofísicos da província Uranífera de Lagoa Real (BA), adquiridos no Projeto São Timóteo. O propósito é contribuir com novas informações que permitem uma caracterização melhor das estruturas geológicas em subsuperfície, associadas ás zonas de mineralização de urânio nesta região. Os dados aeromagnéticos foram reprocessados, focalizando-se no uso criterioso dos procedimentos de correção e técnicas de interpretação. Emprego de técnicas tais como a derivada vertical, sinal analítico e deconvolução de Euler permitiram identificação de diversos lineamentos e feições estruturais na área de estudo. Destacam-se, neste contexto, identificação das estruturas arqueadas na parte centro-oeste da área de estudo, feições circulares nas partes sul e sudeste e diversos lineamentos magnéticos na parte norte. Há indícios que as mudanças nas direções dos lineamentos magnéticos estejam relacionadas com as zonas de fraturas e de cisalhamento e relacionadas com os locais de mineralização de urânio. A grande parte dos corpos com contrastes magnéticos se encontra em profundidades menores que 500 m, mas também há um numero significativo de corpos em profundidades de até 1500 m. Os resultados obtidos também permitiram delimitação de diversas feições estruturais, não identificadas nos levantamentos geológicos locais e nas análises anteriores de dados aerogeofísicos da área de estudo. Progressos também foram alcançados nas análises integradas das informações gama - espectrométricas, permitindo a identificação dos locais afetados por processos de metassomatismo e suas interrelações com as zonas de mineralização de urânio. Os resultados obtidos estão em boa concordância com o arcabouço estrutural da região inferido com base em estudos geológicos, mas significativamente diferentes daqueles obtidos em estudos de dados aerogeofísicos anteriores.
**Palavras-chave:** Distrito Uranífero de Lagoa Real, Projeto São Timóteo, levantamentos aerogeofísicos, derivada vertical, sinal analítico, deconvolução de Euler e estruturas geológicas em subsuperfície.

**ABSTRACT** – *S.N.P. Guimarães & V.M. Hamza - Advances in the characterization of subsurface geological structure of the province uranic Lagoa Real (BA) from aerogeophysical data*. In this work we present results of a recent re-evaluation of airborne magnetometric and gamma spectrometric data of the uranium province of Lagoa Real (BA), acquired under the project São Timoteo. The main purpose is to provide new insights into the magnetic and radiometric characteristics of the structural units in this region, derived from a careful analysis of the aeromagnetic and radiometric data of the study area. We provide details of the correction and interpretation procedures employed in data processing. A remarkable feature of the present work has been the judicious use of such interpretation techniques as the vertical derivative, analytic signal and Euler de-convolution, which have lead to identification of a large number of lineaments and basement features in the study area, not identified in previous studies. Prominent among these are the NW-SE trending lineaments in the northern parts, arc shaped features in the west-central parts and circular features in the south and south-western parts of the Lagoa Real province. There are indications that changes in direction of the lineaments are associated with fracture zones, associated with Uranium mineralization processes. Most of the bodies with magnetic contrasts are located at depths less than 500 meters, but there also a number of bodies at depths extending up to depths of 1500m. The results have also allowed determination of a number of structural features not identified in geologic studies as well as in previous interpretations of aeromagnetic data. Joint interpretation of magnetic and gamma spectrometric data have allowed identification of uranium mineralization zones, associated with local metasomatic processes. We conclude that the results obtained are in good agreement with the local structural framework inferred from geologic studies, but significantly different from those reported in earlier studies.
**Keywords:** Uranium province of Lagoa Real, São Timoteo Project, aero-geophysical data, vertical

# INTRODUÇÃO

A Província Uranífera de Lagoa Real, localizada na região centro-sul do estado da Bahia, constitu o principal local de extração do minério de urânio, atualmente em atividade na América do Sul. As zonas de mineralização nesta província foram descobertas após levantamentos aerogeofísicos realizados no Quadrilátero Ferrífero. Destes, o aerolevantamento da radiação gama realizado durante a execução do projeto São Timóteo em 1979, confirmou a existência de vários bolsões de mineralização de urânio (Oliveira et al., 1985). No entanto, a presença de coberturas lateríticas espessas tornou-se obstáculos impedindo avanços rápidos no emprego de métodos de prospecção geológica convencional e na delimitação das zonas de mineralização. As tentativas iniciais de utilizar resultados de levantamentos aeromagnéticos, com base em técnicas de análises convencionais, não levaram resultados positivos. O estudo inicial da Companhia de Pesquisas de Recursos Minerais (CPRM) se limitou à elaboração dos mapas do campo magnético total, após a incorporação das correções de operações técnicas (CPRM, 1995) e não foram utilizadas as técnicas modernas de interpretação para a identificação das anomalias residuais. O estudo posterior de Pascholati et al. (2003) refere-se textualmente ao uso das técnicas de interpretação baseadas em derivadas espaciais e transposições das anomalias, mas não apresenta detalhes dos procedimentos adotados, adicionalmente encontram-se ausentes neste trabalho as informações importantes sobre as etapas intermediárias de análise e de interpretação, o que torna difícil a avaliação do significado geológico dos resultados apresentados. De fato os conjuntos de lineamentos e feições estruturais deduzidos por Pascholati et al. (2003) apresentam baixo grau de compatibilidade com o arcabouço estrutural inferida a partir das evidências geológicas.

Neste contexto, o propósito do presente trabalho é contribuir com uma reavaliação detalhada dos dados do levantamento aeromagnético, realizado no Projeto São Timóteo. Focalizando-se no uso criterioso das correções (nivelamento e micronivelamento, eliminação dos efeitos de variação diurna e do campo interno) e emprego de técnicas avançadas de interpretação (transposição das anomalias, derivada vertical, sinal analítico e deconvolução de Euler) para identificar as estruturas geológicas em subsuperfície. São apresentadas as sínteses das etapas das correções, do processamento dos dados (técnicas de homogeneização e de filtragens) e de interpretação, junto com mapas que ilustram as anomalias residuais dos magnéticos e radiométricos da Província Uranífera de Lagoa Real. Abordam-se também os significados geológicos dos resultados alcançados através das análises das correlações com as características das zonas de mineralização de urânio.

# CARACTERÍSTICAS GEOLÓGICAS LOCAIS

O contexto geológico e tectônico da região de Lagoa Real faz parte da evolução do Cráton São Francisco e dos ciclos e eventos geológicos sucessivos: Jequié, Transamazônico, Espinhaço e Brasiliano (Almeida 1977; Cordani & Brito Neves 1982; Fyfe 1979), apresenta-se na Figura 1 o mapa geológico simplificado da área de estudo. As unidades litoestratigráficas mais antigas desta região incluem gnaisses e granodioritos do Complexo Gavião e Complexo Paramirim e da faixa Ibitira – Ubiraçaba, de idade Arqueana. Nestas unidades ainda incluem granitos e granitóides anorogênicos calcio-alcalinos quartzitos e rochas metassedimentares da formação Arraias, todas de idade Proterozóica. As unidades mais recentes do embasamento são arenitos e siltitos do Grupo Chapada Diamantina, de idade Mesoproterozóico. Grande parte deste embasamento se encontra cobertos por camadas finas de detrito lateríticos e depósitos aluvionares do período Cenozóico.

O embasamento metamórfico na área de estudo, denominado de complexo granítico-gnáissico de Lagoa Real (faixa central do mapa geológico, Figura 1) é a principal área de interesse do estudo. As feições estruturais identificados na área de estudo incluem as zonas de albititos em forma de arcos (Maruejol et al., 1987) e zonas de cisalhamento (Costa et al., 1985 e Osako, 1999). Conforme indicado no mapa da Figura 2A, grande parte das anomalias radiométricas estão hospedadas nos abititos. Esta faixa estende-se aproximadamente na direção norte-sul e abrange uma área superior a 2.000 km². Nesta região, há indícios de que ocorreram deformações com transporte tectônico de leste para oeste (Figura 2B). Essas deformações foram interpretadas como resultantes de um processo tectônico de colisão N-S, durante o evento Brasiliano (Fuzikawa et al., 1990; Villaça e Hashizyme, 1982; Lobato e Fyfe, 1990). As estruturas do embasamento se encontram arqueadas na parte leste da região central da área de estudo. No entanto, a deformação não atingiu de forma significativa a parte nordeste, onde

o embasamento é caracterizado por estruturas quase lineares de direção noroeste – sudeste. Na parte sudeste da área, o embasamento arqueano possui formato quase circular, sendo limitado nas suas bordas por rochas das fácies xisto verde da faixa Ibitira – Ubiraçaba.

As zonas de cisalhamento são na maioria relacionadas à tectônica de colisão que se desenvolveu durante o evento Brasiliano. Neste evento as rochas do embasamento arqueano de Lagoa Real foram sobreposto às rochas metassedimentares do Supergrupo Espinhaço (Caby e Arthaud, 1987). Retrometamorfismo e alteração metassomática ao longo dessas zonas de cisalhamento possibilitaram a formação de albititos alongados nesta região. Também é comum observar albititos em forma de arcos na parte central da faixa N-S. Este arqueamento é denominado de Torção tipo Helicoidal por Oliveira et al. (1985). As mineralizações de urânio nesta área estão associadas aos albititos e às zonas de cisalhamento, porém nem todos os albititos são mineralizados (Maruejol et al., 1987; Osako,1999).

Segundo Pascholati et al. (2003), as feições identificadas na imagem LANDSAT permitiram delimitação de cinco unidades morfoestruturais, relacionadas com as zonas de mineralização de urânio. Contudo, as feições deduzidas a partir da imagem LANDSAT não apresentam compatibilidade com os lineamentos magnéticos deduzidos a partir dos resultados apresentados pelos mesmos autores. É possível que as feições na imagem LANDSAT, não estejam relacionadas com contrastes nas propriedades magnéticas do embasamento.

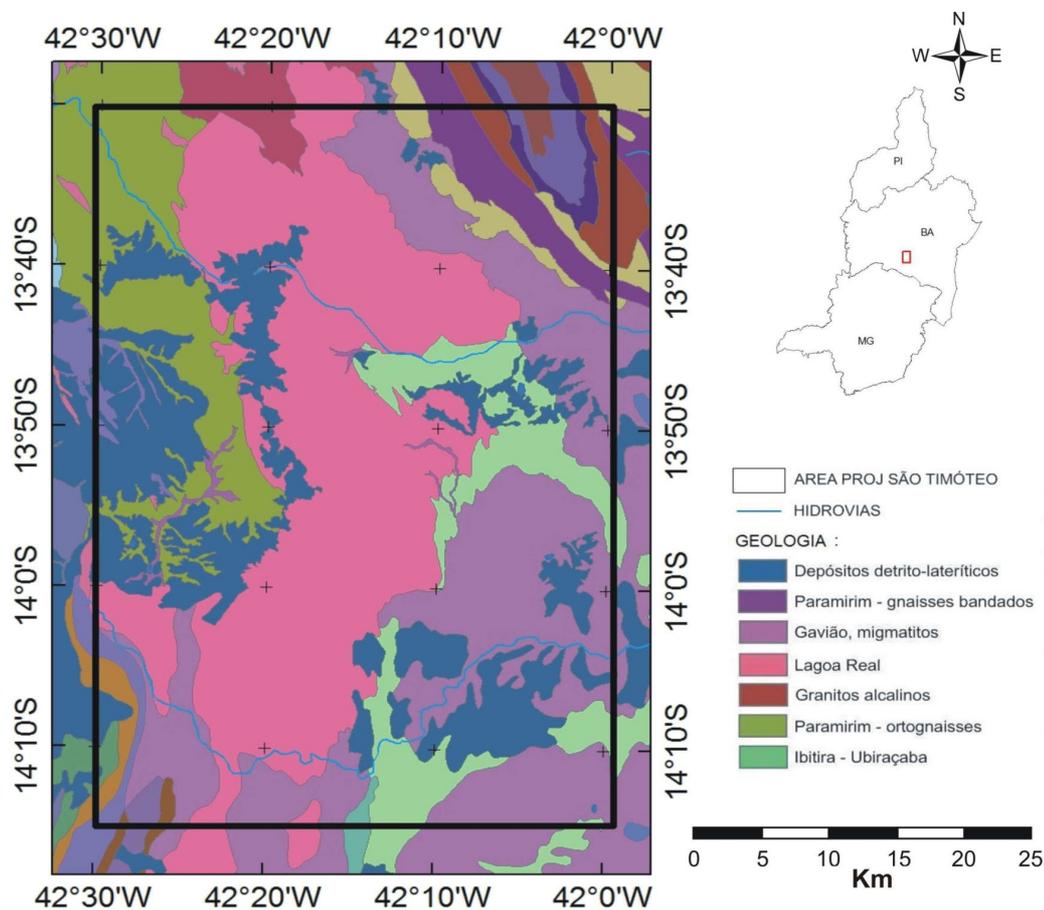

**FIGURA 1.** Mapa geológico simplificado da área do projeto São Timóteo, situada na Província Uranífera Lagoa Real (BA).

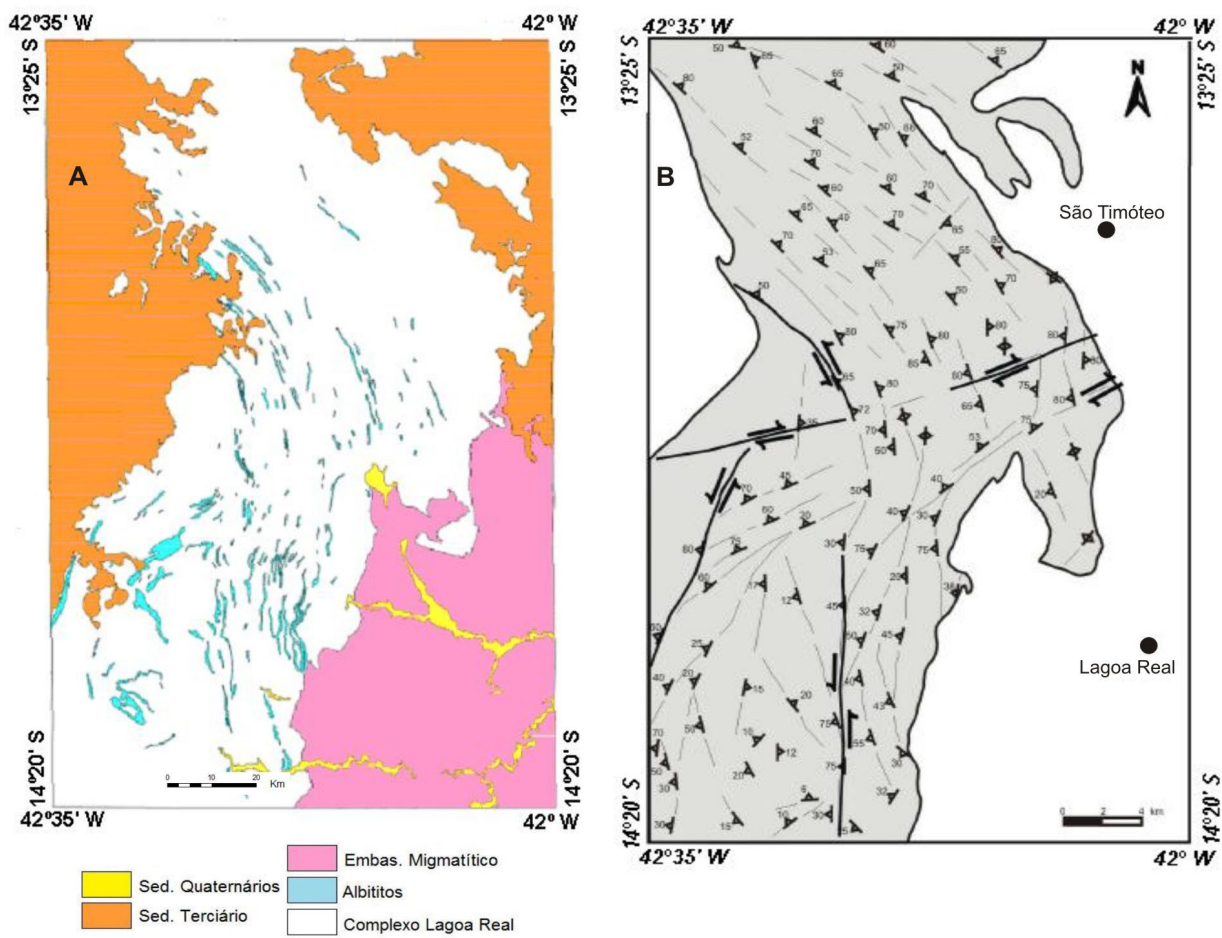

**FIGURA 2.** Feições estruturais identificados na área de estudo. **(A)** Mapa simplificado das zonas de albititos em forma de arcos (Maruejol et al., 1987). **(B)** Linhas de forma inferida com base em dados de direção de foliação e as zonas de cisalhamento (Costa et al., 1985 e Osako, 1999).

## MATERIAIS E MÉTODOS

### BASE DE DADOS

Os dados do Projeto São Timóteo foram disponibilizados para fins de pesquisas acadêmicas pelo CPRM a partir do ano 2000, e foram utilizados em diversos estudos (e.g. Pascholati et al., 2003; Santos, 2006). Os levantamentos aerogeofísicos neste projeto compreenderam linhas de vôo na direção leste-oeste, com espaçamento de 500 m e altura nominal de vôo de 150 m. Os dados foram coletados ao longo de linhas aproximadamente paralelas (E-W) denominadas: linhas de produção. Também foram coletados dados ao longo de algumas linhas perpendiculares (N-S) denominadas: linhas de controle, sendo que os espaçamentos das linhas de controle são cerca de dez vezes maior que as das linhas de produção. O levantamento constou de medições de radiação gama e do campo magnético, utilizando sensores aerotransportados. Os dados magnéticos primários referem-se à intensidade do campo magnético total, expressos em unidades de nano tesla (nT). Os dados gamaespectrométricos referem-se às medições da intensidade de radiação gama emitidas pelos elementos potássio (K), urânio (U) e tório (Th), sendo que os respectivos dados se encontram registrados nos canais correspondentes a esses elementos.

Os dados coletados neste aerolevantamento, realizado na década de 1970, apresentaram vários saltos nas medidas do campo magnético. Há indícios de que estes saltos são artefatos, basicamente devido à ausência de um mecanismo de localização mais confiável, tais como o GPS para navegação e posicionamento da aeronave. Na época do aerolevantamento, o sistema de localização geográfica da aeronave se baseava em cartas de navegação aérea e sistemas físicos baseados no efeito Doppler. Utilizavam-se ainda objetos naturais como referências (e.g. árvore, rios), o que podem ser considerados como possíveis fontes de erros nos dados coletados. Contudo, os problemas desta natureza teriam conseqüências somente na identificação de anomalias em escala local e suas conseqüências podem ser minimizadas através das técnicas de pré-processamento. Os dados aéreos brutos apresentaram informações associadas a ruídos e outras distorções que somente após o devido tratamento das correções, passaram a ter significado físico.

## PROCESSAMENTO DE DADOS

O processamento dos dados aeromagnéticos e aerogamaespectrométricos foi efetuado em etapas distintas. Na etapa inicial de pré-processamento, os dados aeromagnéticos foram corrigidos dos efeitos das variações de altura dos vôos, variação diurna, campo de referência internacional (IGRF), ocorrências de picos (*spikes*), correções de *heading* e de "LAG" (erro sistemático causado pela distancia entre os sensores de medição e o sensor de posicionamento). Os métodos utilizados nessas correções são semelhantes aos aqueles adotados na literatura (ver por exemplo, Hood & Ward, 1969). Na etapa seguinte foram efetuados os procedimentos de nivelamento com a finalidade de introduzir correções nas medidas realizadas em linhas de produção, tomando como referência os padrões observados nas medidas realizadas em linhas de controle (cruzamento). Ainda foram efetuados alguns ajustes em cada cruzamento das linhas, com eliminação de dados considerados como ruidosos, a fim de eliminar informações consideradas tendenciosas, ou seja, identificadas pela presença de feições que seguem as orientações das linhas de vôo.

A incorporação das correções das operações técnicas, de nivelamento e de micro-nivelamento é a etapa necessária para determinações das características do campo magnético regional. Como o nivelamento convencional dos dados magnéticos usando linhas de controle ainda está sujeito a perturbações de diversas origens, adotou-se o procedimento de micro-nivelamento sugerido inicialmente por Minty (1991). Após as etapas de nivelamento e micro-nivelamento, os dados foram processados com a finalidade de montagem de grades regulares. Nesta etapa, foram efetuadas as interpolações dos dados pelo método de Mínima Curvatura, conforme procedimentos sugeridos por Biggs (1974). No presente caso, o tamanho da célula utilizado nas interpolações é de 125 m visto que o espaçamento das linhas de vôo do aerolevantamento foi de 500 m. Esta escolha do tamanho da célula obedece ao critério de Nyquist. Análise comparativa dos campos magnéticos da área de estudo ilustra as melhorias obtidas no pré-processamento de dados. Por exemplo, o campo magnético bruto do setor centro-sul da área de estudo revela a presença de feições lineares nas direções leste-oeste e norte-sul, que são oriundos das operações técnicas na aquisição de dados (Figura 3A). Os resultados obtidos após as devidas correções indicam reduções significativas dos efeitos dessas perturbações (Figura 3B).

No presente trabalho, o pacote de programas computacionais disponíveis em Geosoft® – Oásis Montag (2007) foi utilizado nas tarefas de processamento, análise e de interpretação. Uma das etapas iniciais de interpretação é introduzir correções para a transformação das anomalias visando eliminar eventuais perturbações oriundas das mudanças na direção de magnetização induzida. Foram consideradas as duas formas de transformação, conhecidas como Redução ao Pólo e Redução ao Equador. É importante notar que a transformação das anomalias através de redução ao Equador não necessariamente representa melhorias na análise de dados. Segundo Nabighian et al. (2005) os métodos de redução ao Pólo e ao Equador não apresentam resultados coerentes nas baixas latitudes, sendo que os efeitos variáveis de magnetização remanescente das rochas na área de estudo interferem nos resultados. Para os dados utilizados neste trabalho, a redução ao Pólo foi substituída pela redução ao Equador, visto que a inclinação magnética da área de estudo é baixa. No presente caso, os valores atribuídos aos ângulos de inclinação e declinação do campo são -15° e -20°, respectivamente.

A técnica de derivada espacial (ver, por exemplo, Gunn, 1975) foi utilizada para realçar as altas freqüências e atenuar as baixas. É fisicamente equivalente à medida do campo magnético de dois pontos muito próximos, sendo que o calculo é realizado com base na diferença entre os dados próximos e dividindo o resultado pela separação vertical entre os pontos. Quando aplicadas aos dados magnéticos do campo total anômalo, as derivadas ressaltam as respostas magnéticas dos corpos geológicos mais rasos tais como lineamentos em detrimento aos dos mais profundos. A técnica auxilia também na separação das curvas de anomalias que estejam superpostas lateralmente.

Na maioria dos casos, as feições resultantes da aplicação deste método aparecem como 'cristas', cuja distribuição geográfica parece ocorrer ao longo de linhas, constituindo-se assim chamados de *lineamentos magnéticos*. Considera-se que os mecanismos responsáveis pela conectividade dos lineamentos magnéticos são as mudanças nas propriedades magnéticas e comportamento rúptil das formações geológicas decorrente da atuação de forças tectônicas locais (Costa et al, 1985 e Osako, 1999). No primeiro caso as alterações na direção de lineamentos representam as mudanças nos litotipos. No segundo caso, os locais das mudanças na direção de lineamentos representam as zonas de fraturas locais.

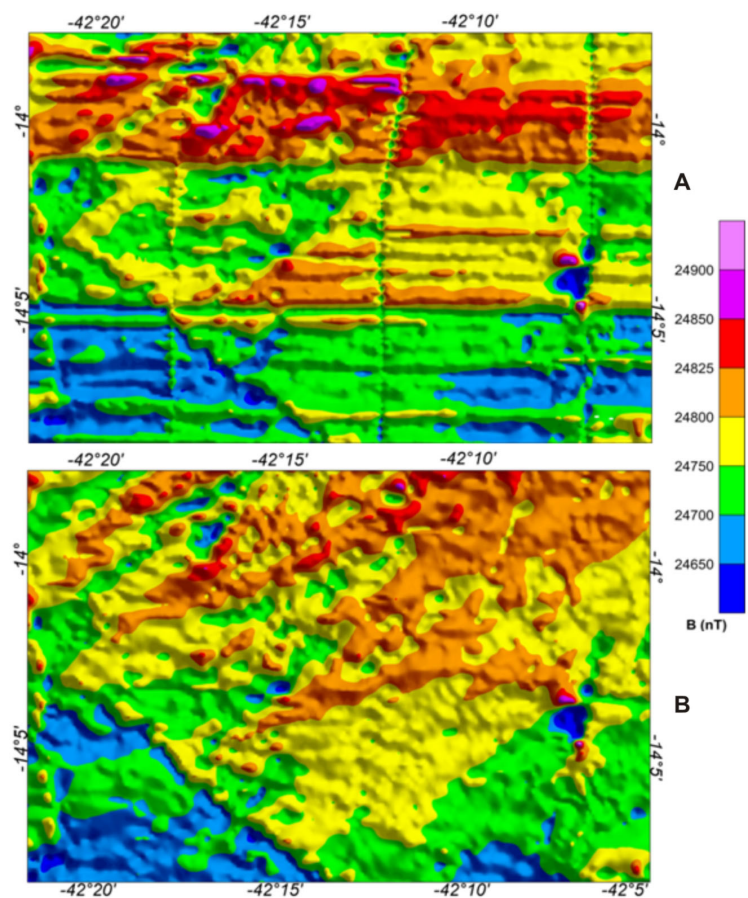

**FIGURA 3.** Exemplo ilustrativo dos resultados do nivelamento e micro nivelamento em dados magnéticos da área selecionada. **(A)** Campo magnético bruto indicando presença de feições lineares que são artifícios gerados no processo técnico de aquisição dos dados. **(B)** Campo magnético micro nivelado onde essas feições foram eliminadas.

Outra técnica utilizada na interpretação foi o Sinal Analítico (Nabighian, 1972; Thompson 1982; Blakely e Simpson, 1986), que basicamente é o módulo da segunda derivada nas três direções do campo magnético, caracterizando-se como um pacote de energia. Na prática, o sinal analítico é considerado como a melhor ferramenta para a localização das bordas de corpos que possuem contraste magnético, trazendo informações relevantes sobre a sua geometria. Quando aplicadas aos dados magnéticos do campo total anômalo, as respostas de sinal analítico ressaltam os limites na superfície dos corpos geológicos com contrastes nas suas propriedades magnéticas.

A técnica de deconvolução de Euler (Nettelton e Cannon, 1962; Clark, 1997; Reid et al., 1990) foi utilizada no presente trabalho para extrair informações sobre a profundidade das fontes magnéticas. Seu resultado independe da direção e da inclinação do campo magnético principal e da orientação das fontes magnéticas. Desta forma o método é relativamente insensível a pequenas distorções do campo. Conforme práticas adotadas na literatura as variações nos graus de homogeneidade do campo são expressas como *índices estruturais*, que especifica o tipo da fonte magnética. De acordo com Thompson (1982) o índice estrutural 0 (zero) representa contatos de diferentes tipos de rochas. Os índices estruturais 0,5 (meio) e 1 (um) representam respectivamente as feições lineares tais como falhas e diques. Os índices 2 (dois) e 3 (três) representam estruturas tubulares e esféricos respectivamente. O sucesso do método é avaliado de forma qualitativa, pelo acumulo de soluções.

No caso de dados gamaespectrométricos os métodos de pré-processamento adotados incluíram correções de altura, do efeito Compton e do ambiente (*background*). A técnica de micro-nivelamento foi a mesma utilizada na análise de dados magnéticos, proposta por Minty (1991). Este procedimento elimina as distorções das linhas de vôo. O processamento realizado envolveu a análise estatística para verificação da consistência dos dados. Como no caso de dados aeromagnéticos a interpolação dos dados foi efetuada pelo método de curvatura mínima (Biggs, 1974) com células de dimensão de um quarto do tamanho do espaçamento do aerolevantamento que, neste caso, foi de 125 m.

# RESULTADOS OBTIDOS

## CAMPO MAGNÉTICO REGIONAL

O nivelamento dos dados primários foi realizado em toda a área de levantamentos aerogeofísicos. Os resultados obtidos nesta etapa permitiram identificação de características magnéticas regionais e elaboração do mapa do campo magnético da Província Uranífera Lagoa Real (Figura 4). A feição marcante neste mapa é a presença de uma região extensa na parte norte da Província onde o campo magnético alcança valores relativamente elevados, na faixa de 24850 à 25000 nT. As extensões laterais ao norte e ao sul desta região é limitada por zonas onde o campo magnético apresenta valores menores que 24700 nT. Os litotipos predominantes que ocorrem nas regiões ao norte e ao sul são as rochas metamórficas do complexo Paramirim de idade Mesoarqueano. Ao sul desta região o campo magnético apresenta valores relativamente menores, predominantemente na faixa de 24700 à 24850 nT. Esta região corresponde a zonas de ocorrência de albititos com granada, anfibolitos, piroxênio, biotita e ortognaisses porfiroclástico, que formam a suíte intrusiva Lagoa Real. O extremo sul da área de estudo é caracterizada por campo magnético com valores no intervalo de 24600 à 24700 nT, o que corresponde a área de ocorrência de ortognaisse migmatítico, com enclaves máficos e ultramáficos do complexo Gavião. Há indícios de que ambas as regiões são constituídos de blocos estruturais segmentados. O mecanismo responsável pelas variações desta natureza no campo magnético total é desconhecido. Uma das possibilidades é ocorrência de diferenças sistemáticas nas propriedades magnéticas e na magnetização remanescente das formações geológicas que compõem o embasamento na área de estudo.

## LINEAMENTOS MAGNÉTICOS

As delimitações dos corpos com contrastes nas propriedades magnéticas na Província Uranífera Lagoa Real foram efetuadas com base nos resultados das derivadas espaciais do campo magnético local. Constam nos Quadros 1A e 1B as características principais dos 28 lineamentos magnéticos identificados na Província Uranífera Lagoa Real. Estes lineamentos apresentaram variabilidades significativas, mas os comprimentos são comparáveis as dimensões espaciais das unidades litológicas locais. Outro fato marcante é a natureza da conectividade entre os lineamentos adjacentes com orientações distintas, que leva a formação de 'arcos' de grande extensão. Os exemplos marcantes são os arcos de lineamentos no setor

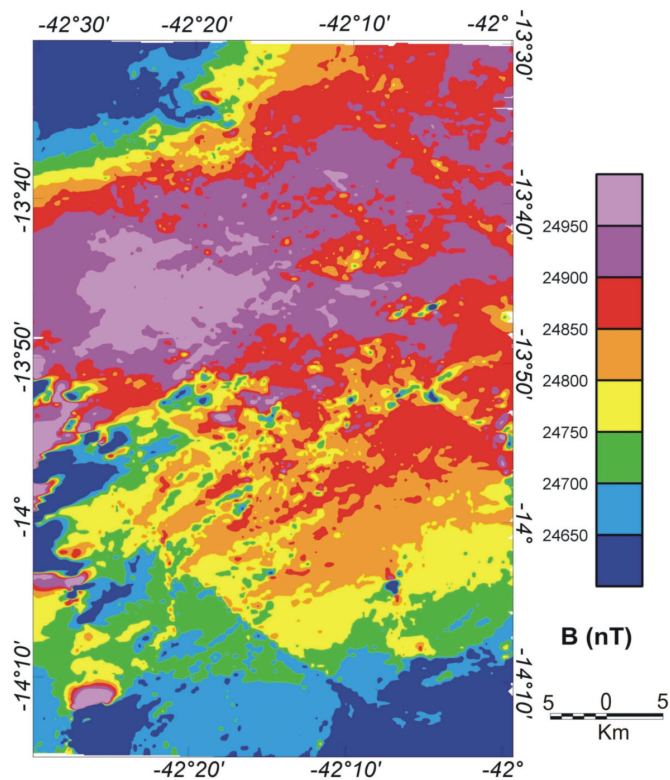

**FIGURA 4.** Mapa do campo magnético corrigido e micro-nivelado da Província Uranífera Lagoa Real.

**QUADRO (1A).** Características dos lineamentos magnéticos identificados L1 à L14.

| Ident. | Coordenadas Sul e Oeste | | Comprimento (km) | Contexto Geológico |
|---|---|---|---|---|
| | Ponto Inicial | Ponto Final | | |
| L1 | 13° 41' 25.46"<br>42° 0' 5.85" | 13° 29' 43.39"<br>42° 8' 16.91" | 27,76 | Formação Rio dos Remédios |
| L2 | 13° 42' 43.02'<br>42° 5' 25.38" | 13° 29' 52.76"<br>42° 16' 49.24" | 32,91 | Complexo Gavião |
| L3 | 13° 44' 52.54'<br>42° 6' 5.28" | 13° 30' 10.15"<br>42° 20' 53.97" | 42,63 | Suíte intrusiva Lagoa Real |
| L4 | 13° 47' 1.13"<br>42° 0' 20.44" | 13° 30' 37.48"<br>42° 23' 20.64" | 51,13 | Complexo Gavião |
| L5 | 13° 52' 33.41"<br>42° 01' 1.47" | 13° 30' 25.99"<br>42° 28' 37.33" | 56,69 | Suíte intrusiva Lagoa Real |
| L6 | 13° 52' 17.55"<br>42° 9' 46.19" | 13° 39' 53.74"<br>42° 28' 20.01" | 65,96 | Suíte intrusiva Lagoa Real |
| L7 | 13° 40' 59.94"<br>42° 5' 37.97" | 13° 54' 42.45"<br>42° 11' 22.55" | 24,58 | Suíte intrusiva Lagoa Real |
| L8 | 13° 49' 54.13'<br>42° 21' 18.4' | 13° 5' 50.59"<br>42° 27' 56.16" | 33,56 | Complexo Paramirim |
| L9 | 13° 56' 48.52"<br>42° 12' 25.2" | 13° 45' 29.60"<br>42° 22' 51.97" | 29,89 | Suíte intrusiva Lagoa Real |
| L10 | 13° 47' 35.15"<br>42° 23' 5.66" | 14° 10' 6.40"<br>42° 29' 6.39" | 68,30 | Lateritos e Complexo Paramirim |
| L11 | 14° 2' 52.31"<br>42° 16' 58.6" | 13° 51' 40.99"<br>42° 21' 39.48" | 26,18 | Cobertura detrito - lateríticas |
| L12 | 13° 53' 51.60"<br>42° 23' 5.25" | 14° 2' 18.26"<br>42° 27' 13.85" | 34,45 | Cobertura detrito - lateríticas |
| L13 | 14° 4' 26.70"<br>42° 25' 14.7' | 13° 52' 15.91"<br>42° 29' 59.77" | 33,50 | Complexo Paramirim |
| L14 | 13° 50' 21.86"<br>42° 19' 39.0" | 14° 0' 27.92"<br>42° 28' 34.77" | 25,20 | Complexo Paramirim |

sudoeste da província e a feição quase circular no setor sudeste. Essas feições podem estar refletindo o arcabouço estrutural desta região, particularmente as ocorrências de zonas de empurrão, com vergência de oeste para leste (Figura 5A). No setor norte os lineamentos possuem direções predominantes de NW – SE, coincidentes com as direções preferenciais identificados nos levantamentos geológicos (Figura 2B). No setor sul há um lineamento magnético de grande extensão na direção NW-SE, que não consta nos mapas geológicos da província (Figura 5B). As características das anomalias magnéticas associadas apontam para um dique de rochas básicas de grande extensão, inseridas no complexo metamórfico desta região, no período Fanerozóico.

## SINAL ANALÍTICO E VARIAÇÕES NAS PROPRIEDADES MAGNÉTICAS

O mapa do sinal analítico da Província Uranífera Lagoa Real foi elaborado com a finalidade de delimitar os corpos que possuem contrastes magnéticos. De acordo com os resultados obtidos as magnitudes do sinal analítico variam de -0,03 a +0,15 nT/m na área de estudo. A variabilidade na amplitude do sinal analítico é significativa, mas é possível identificar áreas onde os valores médios são relativamente uniformes (Figura 6). Contudo, o fato notável é que os formatos espaciais dessas regiões de amplitudes uniformes apresentam semelhanças marcantes com as áreas delimitadas pelos conjuntos de lineamentos magnéticos. Parece que as áreas de sinal analítico uniforme, limitadas pelos

**QUADRO (1B).** Características dos lineamentos magnéticos identificados L15 à L28.

| Ident. | Coordenadas Sul e Oeste | | Comprimento (km) | Contexto Geológico |
|---|---|---|---|---|
| | Ponto Inicial | Ponto Final | | |
| L15 | 14° 2' 43.54"<br>42° 26' 12.8" | 14° 4' 58.35"<br>42° 29' 47.74" | 8,98 | Super grupo Espinhaço |
| L16 | 14° 14' 38.90"<br>42° 4' 52.23" | 14° 0' 9.26"<br>42° 22' 34.43" | 41,78 | Suíte intrusiva Lagoa Real |
| L17 | 14° 2' 36.64"<br>42° 21' 16.84" | 14° 8' 5.95"<br>42° 20' 31.28" | 11,15 | Complexo Gavião |
| L18 | 14° 14' 39.15"<br>42° 15' 38.82" | 14° 9' 10.77"<br>42° 23' 17.61" | 17,81 | Complexo Gavião |
| L19 | 13° 59' 22.88"<br>42° 8' 47.75" | 14° 14' 3.49"<br>42° 12' 20.88" | 30,68 | Cobertura detrito - lateríticas |
| L20 | 13° 50' 19.05"<br>41° 59' 37.07" | 13° 51' 51.63"<br>42° 4' 39.22" | 11,38 | Xisto verde Ibira-Ubiraçaba |
| L21 | 14° 6' 14.83"<br>41° 59' 36.04" | 13° 54' 42.58"<br>42° 0' 19.18" | 69,57 | Complexo Gavião |
| L22 | 13° 58' 17.16"<br>42° 4 57.14`` | 14° 3' 15.10"<br>42° 7' 6.20" | 11,53 | Formação Boquira |
| L23 | 14° 2' 0.68"<br>42° 1' 33.20" | 14° 5' 57.33"<br>42° 5' 59.64" | 11,39 | Formação Boquira |
| L24 | 14° 3' 36.12"<br>42° 0' 23.66" | 14° 8' 2.71"<br>42° 5' 12.49" | 12,48 | Xisto verde Ibira-Ubiraçaba |
| L25 | 14° 3' 38.68"<br>42° 3' 52.35" | 14° 10' 43.84"<br>42° 7' 42.32" | 8,42 | Cobertura detrito - lateríticas |
| L26 | 14° 8' 24.23"<br>41° 59' 26.75" | 14° 11' 12.75"<br>42° 6' 41.20" | 1,50 | Greenstones Belts Ibira-Ubiraçaba |
| L27 | 14° 12' 54.49"<br>42° 25' 24.20" | 14° 14' 27.70"<br>42° 26' 16.33" | 3,71 | Suíte intrusiva Lagoa Real |
| L28 | 14° 13' 3.11"<br>42° 27' 17.20" | 14° 14' 3.05"<br>42° 28' 18.18" | 3,26 | Supergrupo Espinhaço |

lineamentos magnéticos, determinam o arcabouço estrutural do embasamento.

### DECONVOLUÇÃO DE EULER E FONTES MAGNÉTICAS

Na utilização da técnica de deconvolução de Euler procurou-se extrair informações sobre a profundidade e as características estruturais das fontes magnéticas na Província Uranífera Lagoa Real. Conforme práticas adotadas na literatura o acúmulo de soluções na aplicação desta técnica foi considerado como indicativo da presença em subsuperfície de corpos com contrastes magnéticos. Ainda, as variações nos graus de homogeneidade do campo, que foram expressas como *índices estruturais* foram utilizados na especificação do tipo da fonte magnética. A distribuição geográfica das fontes magnéticas identificadas desta forma é apresentada no conjunto dos mapas da Figura 7. Nota-se a presença de um número significativo de corpos magnéticos (ou seja, o acúmulo de soluções na deconvolução de Euler) nas partes norte, nordeste, sudoeste e sudeste da área de estudo. Considerou-se que a dispersão nos acúmulos das soluções é ocasionada pelo formato complexo do corpo e pelas heterogeneidades nas suas propriedades magnéticas. As linhas contínuas neste conjunto de mapas representam lineamentos determinadas com base na derivada vertical do campo magnético.

No caso do índice estrutural n=0 (Figura 7A), que corresponde aos contatos de diferentes litotipos presentes no embasamento, os corpos associados possuem profundidades que variam de 0 a 300 m. Referindo-se ao caso de índice estrutural n=0.5 (Figura 7B) os resultados obtidos apontam para correspondências com as principais falhas mapeadas em levantamentos geológicos. As profundidades desses falhamentos inferidas variam de 100 a 500 m. No caso de índice estrutural n=1 (Figura 7C), que corresponde aos diques, encontramos as profundidades no intervalo de 200 a 750 m. Para o índice estrutural n=2 (Figura 7D), que corresponde a estruturas tubulares, as profundidades inferidas variam de 400 a 1200 m. Por fim, para o índice estrutural n=3 (Figura 7E) o resultado das profundidades inferidas variam de 500 a 1500 m.

## Correlações com Anomalias Radiométricas

Avaliações das correlações com as anomalias radiométricas da Província Uranífera Lagoa Real constituiram as atividades da etapa final deste trabalho. A distribuição espacial das anomalias radiométricas (ver o conjunto de mapas da Figura 8) indica que as zonas com maior concentração de radio-elementos se encontram distribuída ao longo de uma faixa central da área de estudo, com direção predominante N-S. A largura desta faixa é variável, sendo maior na parte norte e menor na parte sul. Um fato notável neste contexto é o posicionamento desta faixa em relação ao conjunto lineamentos magnéticos identificado no mapa da Figura (5A). Parece que a extensão lateral desta faixa é limitada pelo conjunto de lineamentos magnéticos. De modo geral, a parte desta faixa onde ocorrem concentrações elevadas de radio-elementos coincide com localização do embasamento metamórfico afetada pela albitização. De acordo com as estimativas preliminares os dados de levantamentos radiométricos indicam valores médios de 7,7 ppm para o urânio (Figura 8A), 18,2 ppm para o tório (Figura 8B) e 1,65 % para o potássio (Figura 8C). A faixa de anomalias radiométricas engloba a maior parte das áreas de exploração do urânio e zonas de mineralizações conhecidas.

Há indícios de que a topografia da região também influi na ocorrência de anomalias radiométricas, sendo que as principais anomalias se encontram em áreas com baixo relevo. Isto acontece, em grande parte, devido à atuação seletiva dos processos de lixiviação, uma vez que grande numero de minerais portadores do urânio e tório se encontra localizado próximo à rede de drenagem.

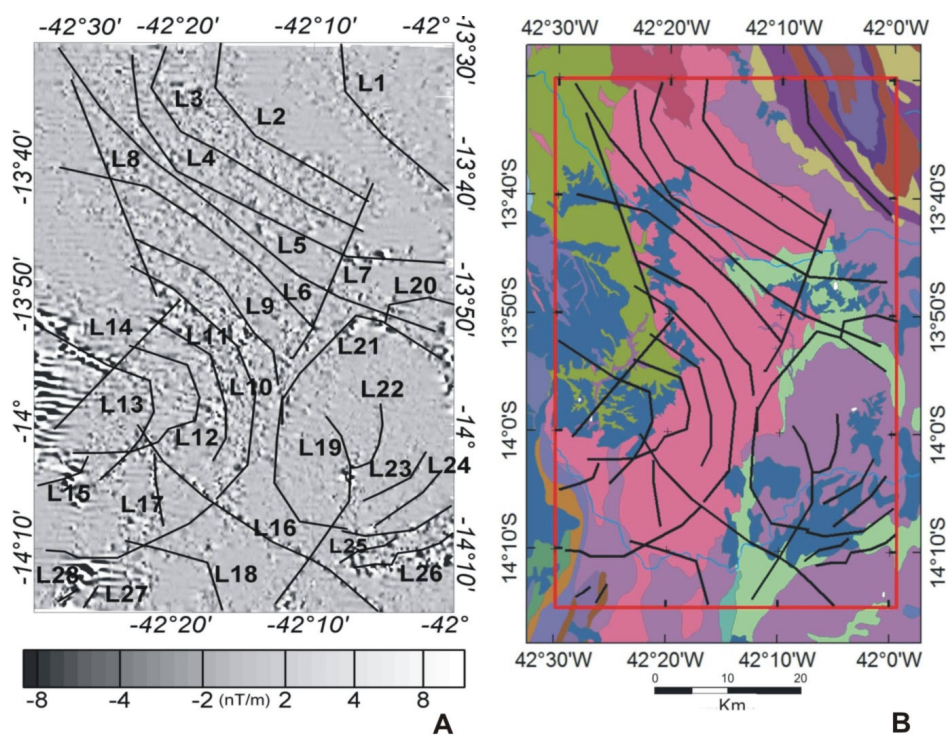

**FIGURA 5.** Mapa da derivada vertical do campo magnético
e a sua correlação com as características geológicas da Província Uranífera Lagoa Real.
**(A)** Conjunto de lineamentos magnéticos identificados na derivada vertical.
**(B)** Superposição dos mapas geológicos e derivada vertical do campo magnético
total ilustrando as correlações entre os lineamentos magnéticos identificados e a geologia da área.

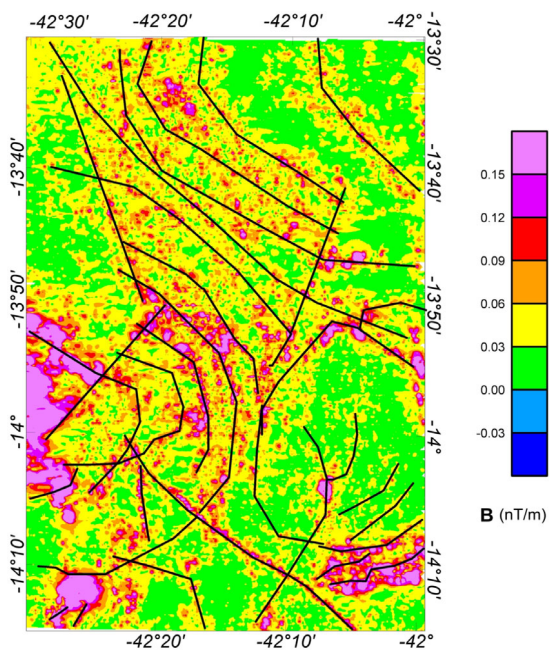

**FIGURA 6.** Mapa do sinal analítico do campo magnético. Nesta figura ilustra-se a distribuição geográfica das amplitudes do sinal analítico junto com a localização de lineamentos magnéticos identificados.

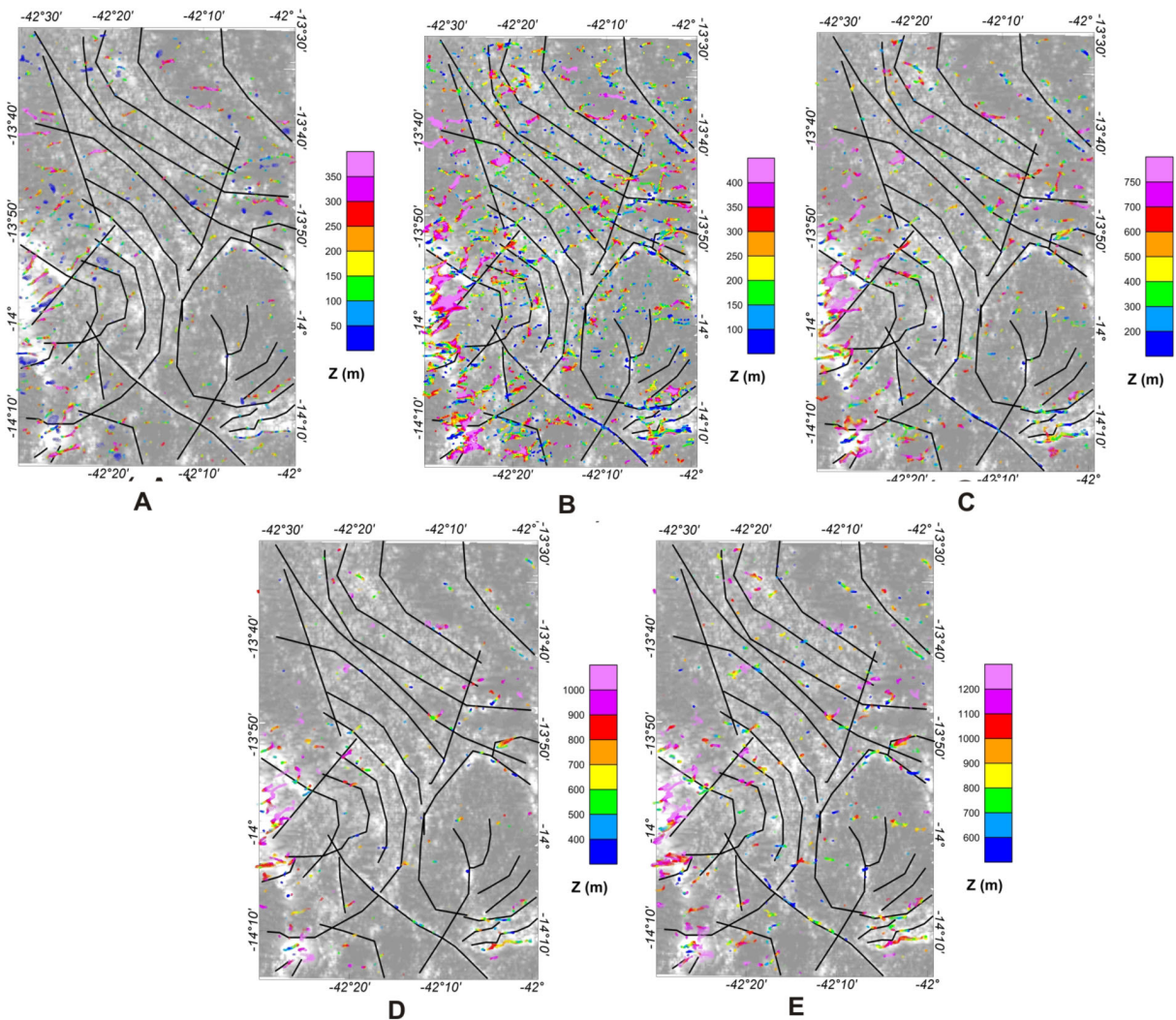

**FIGURA 7.** Resultados do método de Deconvolução de Euler aplicado ao sinal analítico do campo magnético com índices estruturais: **(A)** N = 0 (contatos); **(B)** N = 0.5 (falhas); **(C)** N = 1 (diques); **(D)** N = 2 (cilindros) e **(E)** N = 3 (dipolos). As escalas de cores nessas figuras indicam as profundidades das fontes magnéticas e as linhas contínuas representam lineamentos magnéticos determinados com base na derivada vertical do campo magnético.

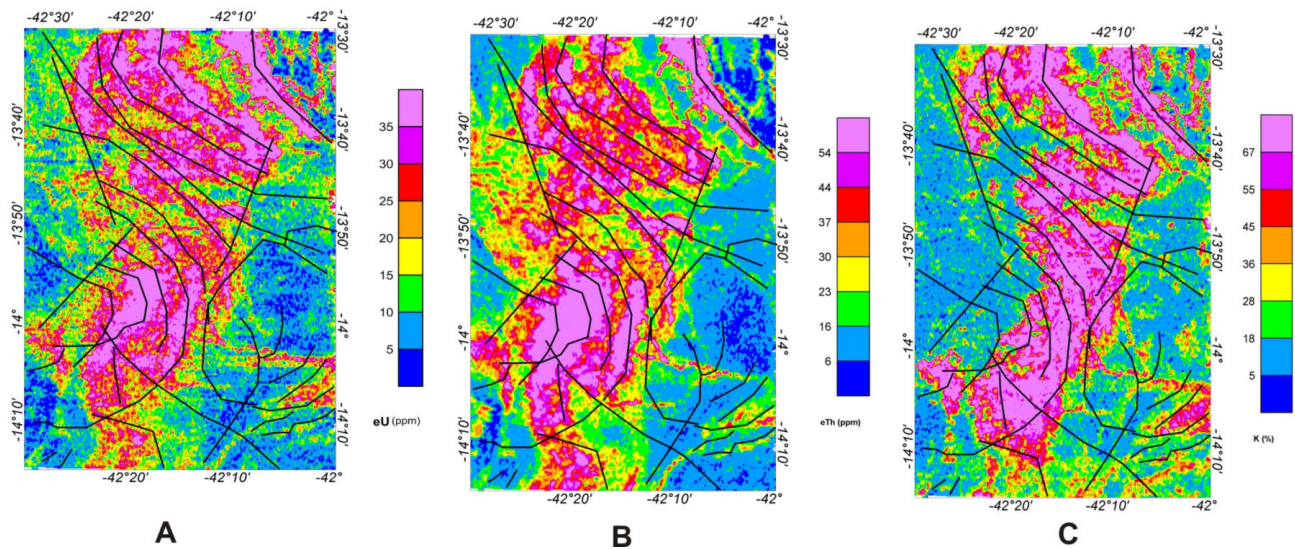

**FIGURA 8.** Mapas de anomalias radiométricas associadas às distribuições espaciais dos elementos radioativos. (A) Urânio, (B) Tório e (C) Potássio.

## CONCLUSÕES

Os resultados obtidos no presente trabalho permitiram avanços importantes na identificação e delimitação das estruturas geológicas em subsuperfície da Província Uranífera Lagoa Real. Destacam-se, neste contexto, as estruturas arqueadas na parte centro-oeste, feições circulares nas partes sul e sudeste e diversos lineamentos magnéticos na parte norte. Há indícios que os locais das mudanças nas direções dos lineamentos magnéticos estão relacionados com as zonas de fraturas e de cisalhamento e são próximos aos locais de mineralização de urânio. A grande parte dos corpos com contrastes magnéticos se encontra em profundidades menores que 500 m, mas também há um numero significativo de corpos em profundidades de até 1500 m. Progressos também foram alcançados nas análises integradas das informações gamaespectrométricas, permitindo a identificação dos locais afetados por processos de metassomatismo e suas interrelações com as zonas de mineralização de urânio. Permitiram ainda, a delimitação de diversas feições estruturais, não identificados nos levantamentos geológicos locais e nas análises anteriores de dados aerogeofísicos.

As conclusões principais desse trabalho são:

Os conjuntos de lineamentos e feições estruturais determinadas no presente trabalho se encontram em boa concordância com o arcabouço estrutural inferida a partir das evidências geológicas (Lobato et al., 1985), mas significativamente diferentes daqueles apresentados no estudo anterior por Pascholati et al., (2003). Destacam-se, neste contexto, as delimitações das estruturas arqueadas na parte centro oeste da área de estudo, das feições circulares nas partes sul e sudeste e dos diversos lineamentos magnéticos na parte norte.

Há indícios que as mudanças nas direções dos lineamentos magnéticos estejam relacionadas com as zonas de fraturas e de cisalhamento e relacionadas com os locais de mineralização de Urânio. A grande parte dos corpos com contrastes magnéticos se encontra em profundidades menores que 500 m, mas também há um número significativo de corpos em profundidades de até 1500 m.

Os resultados obtidos também permitiram delimitação de diversas feições estruturais, não identificados nos levantamentos geológicos locais e nas análises anteriores de dados aerogeofísicos da área de estudo. Progressos também foram alcançados nas análises integradas das informações gamaespectrométricas, permitindo a identificação dos locais afetadas por processos de metassomatismo e suas interrelações com as zonas de mineralização de Urânio.

A redução ao pólo, ou a redução ao equador de dados do campo magnético com pequena inclinação magnética não obtém resultados satisfatórios para localização das fontes magnéticas. A melhor ferramenta utilizada para este fim foi o sinal analítico.

Pudemos fornecer dados de localização e profundidade da anomalia magnética selecionada, através do cruzamento das informações de sinal analítico, deconvolução de Euler, e através da sobreposição ao mapa geológico revelamos a coerência litológica da fonte magnética, classificando satisfatoriamente as informações obtidas.

Num trabalho recente (Guimarães & Hamza,

2009) demonstrou-se que é possível estabelecer correlações significativas entre feições identificados em levantamentos aeromagnéticos e características geológicas da área do projeto São Timóteo. Contudo, os estudos anteriores não alcançaram resultados semelhantes. Por exemplo, os lineamentos magnéticos identificados por Pascholati et al. (2003) não apresentam correlações significativos com as estruturas e feições mapeados nos estudos geológicos. Os fatores decisivos que propiciaram o sucesso do presente trabalho parece ser a execução criteriosa das etapas das correções dos dados primários e das técnicas de análise e de interpretação no processamento de dados aerogeofísicos do projeto São Timóteo.

## AGRADECIMENTOS

## REFERÊNCIAS BIBLIOGRÁFICAS